\documentclass[aps,pra,superscriptaddress,twocolumn,nofootinbib]{revtex4-1}
\usepackage{amsmath}
\usepackage{amssymb}
\usepackage{dsfont}
\usepackage{xcolor}
\makeatletter
\usepackage{amssymb}
\usepackage{graphicx}
\usepackage{epstopdf}
\usepackage{epsfig}
\usepackage{subfigure}
\usepackage{xfrac}
\usepackage[colorlinks,urlcolor=blue,citecolor=blue]{hyperref}
\makeatother

\newcommand{\ket}[1]{|#1\rangle}

\usepackage{cancel,ifthen}
\usepackage[normalem]{ulem}

\newcommand{\bref}[3]{{#1}\href{#2}{#3}}

\newcommand{\comment}[2][NoInPuT]{\ifthenelse{\equal{#1}{NoInPuT}}{}{{\color{blue}\sout{#1}}}{\color{red} #2}}

\begin{document}
\allowdisplaybreaks
\title{Polar core vortex dynamics in disc-trapped homogeneous spin-1 Bose-Einstein condensates}

\author{Matthew Edmonds}
\email{m.edmonds@uq.edu.au}
\affiliation{ARC Centre of Excellence in Future Low-Energy Electronics Technologies, School of Mathematics and Physics, University of Queensland, St Lucia, QLD 4072, Australia}
\affiliation{Department of Physics \& Research and Education Center for Natural Sciences, Keio University, Hiyoshi 4-1-1, Yokohama, Kanagawa 223-8521, Japan}
\author{Lewis A. Williamson}
\affiliation{ARC Centre of Excellence for Engineered Quantum Systems, School of Mathematics and
Physics, University of Queensland, St Lucia, Queensland 4072, Australia}
\author{Matthew J. Davis}
\affiliation{ARC Centre of Excellence in Future Low-Energy Electronics Technologies, School of Mathematics and Physics, University of Queensland, St Lucia, QLD 4072, Australia}

\date{\today{}}

\begin{abstract}\noindent
We study the dynamics of polar core vortices in the easy plane phase of an atomic spin-1 Bose-Einstein condensate confined in a two-dimensional disc potential. A single vortex moves radially outward due to its interaction with background flows that arise from boundary effects. Pairs of opposite sign vortices, which tend to attract, move either radially inward or outward, depending on their strength of attraction relative to boundary effects. Pairs of same sign vortices repel. Spiral vortex dynamics are obtained for same-sign pairs in the presence of a finite axial magnetization. We quantify the dynamics for a range of realistic experimental parameters, finding that the vortex dynamics are accelerated with increasing quadratic Zeeman energy, consistent with existing studies in planar systems.  
\end{abstract}
\maketitle
\section{Introduction}
Quantum gases have emerged over the last three decades as exemplar platforms to probe both the phenomenology of atomic systems and also as a route to study the fundamental physics of many-body quantum mechanical systems in a clean and highly controllable environment \cite{bloch_2008}. Experiments can control the potential landscape, dimensionality and interactions between atoms to realise novel quantum states which may be challenging to access with condensed matter counterparts \cite{vale_2021,baroni_2024}.

Atomic quantum gases forming Bose-Einstein condensates with internal spin degrees of freedom present an opportunity to explore magnetic effects due to their inherent spinor interactions, which at the mean-field level present both antiferromagnetic and ferromagnetic phases \cite{ohmi_1998,kawaguchi_2012}. Experiments have demonstrated optically-confined spinor condensates of $^{23}$Na \cite{stamperkurn_1998,miesner_1999}, $^{87}$Rb \cite{barrett_2001,chang_2004}, $^{52}$Cr \cite{pasquiou_2011} and $^{7}$Li \cite{huh_2021}. The spinor structure of these systems leads to excitations that are topological in nature, which proliferate in space without their characteristics changing, and are robust to external perturbations. Early experimental works have observed the formation of spin domains \cite{stenger_1998}, and the realization of non-singular \cite{leanhardt_2003} and singular \cite{sadler_2006} excitations supported by the magnetic interactions between atoms in the gas. Antiferromagnetic interactions were shown to host stable half-quantum vortices \cite{seo_2015,seo_2016}, as well as skyrmions supported by ferromagnetic interactions \cite{leslie_2009} and vortex lines in a ferromagnetic three-dimensional spinor fluid with SO(3) symmetry \cite{weiss_2019} and the observation and subsequent decay of ferromagnetic singularly quantized excitations \cite{xiao_2021}. Related work has led to the identification of families of solutions classified by their topological properties \cite{kobayashi_2012,eto_2012,borgh_2016}, as well as knot \cite{kawaguchi_2008}, skyrmion \cite{eto_2013a} and instanton \cite{zenesini_2024} excitations.

Spinor condensates are intrinsically sensitive to the local magnetic environment. This allows the spinor system to manifest different phases as the populations of the magnetic sublevels are controlled with an external Zeeman field, which has motivated ongoing theoretical interest into the interplay of nonlinear and spin interactions. The spinor gas can support stable vortices with winding numbers which can be quantized with either singular \cite{lovegrove_2012,lovegrove_2014,borgh_2014} or non-singular windings. This allows access to a broader class of topological phenomena compared to scalar condensates \cite{isoshima_2002}. This includes vortices such as the coreless Mermin-Ho \cite{ho_1998,mizushima_2002} and polar core \cite{saito_2006} vortex which exist in the ferromagnetic `easy-plane' phase. By rotating the spinor gas vortex lattices have been predicted with fractionalized vortices \cite{ji_2008}, while treating the excitation as a point particle yields insight into their superfluid dynamics \cite{turner_2009,williamson_2016}.

Effects which depend on transitions between different equilibrium states play a prominent role in superfluid systems. The addition of magnetic spin degrees of freedom facilitate a richer physical behaviour, due to the larger phase diagram and physical parameter space of the spinor condensates \cite{stamperkurn_2013}.  
Interest in these systems has focused on effects such as the Berezinskii–Kosterlitz–Thouless transition \cite{mukerjee_2006,james_2011,kobayashi_2019,underwood_2023,underwood_2024}, finite-temperature \cite{nikuni_2003,endo_2011,pietila_2010,barnett_2011} and thermalization effects \cite{fujimoto_2019,prufer_2022}, as well as quenches between magnetic phases which can give rise to the Kibble-Zurek scaling of defects \cite{saito_2007,lamacraft_2007}, universal coarsening dynamics \cite{williamson_2016a,williamson_2017,williamson_2019,bourges_2017,symes_2017,fujimoto_2018,schmied_2019,schmied_2019b,prufer_2018,huh_2024},  
spin turbulence \cite{kang_2017} and the possibility of observing rogue waves \cite{siovitz_2023}. 
Adiabatic quenches in a spin-1 $^{87}$Rb condensate across a quantum phase transition have also been studied \cite{hoang_2016}.

Although harmonic confining potentials have provided a great deal of insight into the atomic superfluids, recently planar condensates have been realized using optical box potentials \cite{navon_2021}, a situation allowing a closer comparison with theoretical models. It has been shown experimentally how different trapping configurations can realise one \cite{meyrath_2005,vanes_2010}, two \cite{chomaz_2015} and three-dimensional homogeneous Bose gases \cite{gaunt_2013}, as well as for two and three-dimensional Fermi gases \cite{hueck_2018,mukherjee_2017}. Then, digital micromirror devices (DMDs) allow potentials with dynamical profiles to be realised \cite{gauthier_2016}, yielding a new opportunity to engineer novel superfluid states and investigate the nonequilibrium phenomenology of quantum fluids \cite{gauthier_2019,johnstone_2019}.

Due to their sensitivity to magnetic fields, spinor gases have been proven as efficient magnetometers \cite{vengalattore_2007} with subsequent work demonstrating a spin-echo approach \cite{eto_2013b} and phase-contrast imaging for reduced atomic losses \cite{sekiguchi_2021}. Related to this is the emerging field of atomtronics whose goal is to use quantum mechanical systems to build analogies of traditional electronic devices \cite{amico_2021,amico_2022}. Here spinor gases' additional spin degrees of freedom provide a further opportunity to engineer novel magnetic devices in different dimensional scenarios \cite{mistakidis_2023}, such as interferometers capable of sensitively measuring the local magnetic and rotational environment \cite{helm_2018}.  
 
The internal level structure of spinor systems provides a resource for simulating synthetic forms of matter. One prominent example is the ability of quantum gas systems to host artificial gauge potentials \cite{dalibard_2011,goldman_2014}. Such a situation can be realised using optical Raman couplings between different hyperfine states \cite{spielman_2009}. This has allowed spin-1 spin-orbit couplings with bosons \cite{lan_2014} to be realised \cite{campbell_2016} and explored theoretically \cite{prixley_2016,chen_2020,zhang_2023,gangwar_2024}. The existence of a spin-orbit coupling in a quantum gas has allowed spintronic effects such as the Datta-Das transistor to be suggested \cite{vaishnav_2008} and demonstrated \cite{madasu_2022}. Artificial gauge potentials can also be induced by confining a spinor gas on a cylinder \cite{ho_2015}, while in three dimensions the presence of the artificial gauge potential causes the spinor fluid to manifest a Dirac string \cite{xu_2024}. Finally, following their discovery the existence of stable quantum droplet phases has been subjected to both experimental and theoretical scrutiny. The large parameter space of spinor gases means that the predicted droplet phases \cite{uchino_2010,yogurt_2022,yogurt_2023} contribute new insights into beyond-mean-field effects, such as unusual magnetic vortices in the presence of both dipolar and spinor interactions \cite{li_2024}.

In a ferromagnetic spin-1 condensate the Zeeman fields can be tuned so that the ground-state magnetization is transverse to the applied Zeeman field, with easy-plane symmetry \cite{murata_2007}. In two dimensions this phase supports polar-core spin vortices, which consist of circulation of transverse spin density arising from equal and opposite circulation of two of the spin components, with a core filled by the third component \cite{isoshima_2001}. Polar-core vortices play a fundamental role in both the quench dynamics \cite{sadler_2006,saito_2007,lamacraft_2007,williamson_2016} and equilibrium properties \cite{underwood_2023,tang_2025} of the easy-plane condensate, and exhibit dynamics qualitatively distinct from vortices in scalar condensates \cite{turner_2009}. Studying the dynamics of small numbers of polar-core vortices provides useful insight into the rich behaviour offered by spinor condensates in the easy-plane phase.

The purpose of the present work is to investigate the static and dynamical properties of individual and pairs of polar core vortices confined on a circular disc in the ferromagnetic easy-plane phase of a spin-1 atomic Bose-Einstein condensate. This is distinct from previous studies which have considered harmonic confinement \cite{stevenshough_2024}, or worked with a planar system where angular momentum is not conserved \cite{williamson_2016,williamson_2021}. 

The article is structured as follows -- in Sec.~\ref{sec:model} we detail the appropriate mean-field model describing the magnetic spinor gas. Following this in Sec.~\ref{sec:num} we present our numerical findings exploring the stationary states of pinned polar core vortices as the magnetic Zeeman field is varied, along with the accompanying spinor vortex dynamics. The remainder of Sec.~\ref{sec:num} focuses on understanding the dynamics of pairs of excitations, with opposite windings forming polar core vortex `dipoles' or with same-sign windings. The article concludes with a summary of our findings. 

\section{\label{sec:model}Theoretical background} 
\subsection{Spin-1 model}
\begin{figure}[t]
    \centering
    \includegraphics[width=1.0\columnwidth]{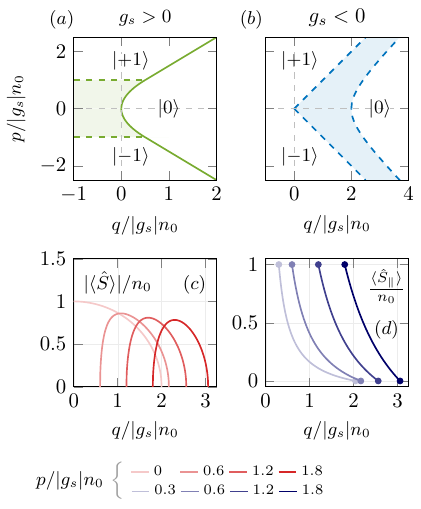}
    \caption{(color online) Ground state phases of Eq.~\eqref{eqn:ham_s} are shown in (a) for (a) antiferromagnetic $g_s>0$ and (b) ferromagnetic $g_s<0$ (b) interactions with $\ket{j}\equiv\ket{m_{\rm F}=j}$. Panel (c) and (d) show examples of the spin densities $|\langle\hat{S}\rangle|/n_0$ (Eq.~\eqref{eqn:den_s}) and $\langle\hat{S}_{\parallel}\rangle/n_0$ (Eq.~\eqref{eqn:den_p}) respectively for fixed values of $p/q_0$. Here $n_0$ defines the mean mass density.}
    \label{fig:gnd}
\end{figure}

We consider a dilute gas of $N$ bosonic atoms of mass $M$ that can occupy the three spin states $m_{\rm F}=-1,0,1$ which are described by the spinor $\boldsymbol{\Psi}({\bf r})=(\psi_{-1}({\bf r}),\psi_{0}({\bf r}),\psi_{+1}({\bf r}))^{T}$. The spinor Hamiltonian is written as
\begin{equation}\label{eqn:ham_s}
    H = \int d^{2}{\bf r}\bigg[\boldsymbol{\Psi}^{\dagger}\hat{H}_{0}\boldsymbol{\Psi}+\frac{g_n}{2}n({\bf r})^2+\frac{g_{s}}{2}|\langle\hat{S}({\bf r})\rangle|^2\bigg].
\end{equation}
Here the quasi-two-dimensional density and spin scattering parameters are defined as $g_n=(c_0+2c_2)/3\sqrt{2\pi l_z}$ and $g_s=(c_2-c_0)/3\sqrt{2\pi l_z}$ and $c_{\rm F}=4\pi\hbar^2 a_F/M$ defines the scattering parameter for pairs of atoms with total spin $F$, while $l_z$ defines the axial length scale. Then the densities for the mass and spin are $n({\bf r})=\boldsymbol{\Psi}^{\dagger}({\bf r})\boldsymbol{\Psi}({\bf r})$ and $\langle\hat{S}_{\mu}({\bf r})\rangle=\boldsymbol{\Psi}^{\dagger}\hat{S}_{\mu}\boldsymbol{\Psi}$ respectively, where $\hat{S}_{\mu}$ is a spin-1 Pauli matrix with $\mu\in\{x,y,z\}$ \cite{pm_def}. The total atom number is $N=\int d{\bf r}\ n({\bf r})$. The single-particle Hamiltonian appearing in Eq.~\eqref{eqn:ham_s} is 
\begin{equation}
    \hat{H}_0 = \bigg(-\frac{\hbar^2}{2M}\nabla^2+U({\bf r})\bigg)\otimes\mathds{1}_{\rm 3\times 3}-p\hat{S}_z+q\hat{S}_{z}^{2},
\end{equation}
here $p$ and $q$ are the magnitudes of the linear and quadratic Zeeman shifts, while $U({\bf r})$ describes the disc potential confining the atoms, defined as 
\begin{equation}\label{eqn:disc}
    U({\bf r})=\frac{U_0}{2}\sum_{j=1}^{2}\bigg[1-(-1)^{j}\tanh\bigg(\frac{r+(-1)^{j}R_0}{\sigma}\bigg)\bigg],
\end{equation}
here $r$ is the two-dimensional radial coordinate while $U_{0}$ defines the depth of the potential, $R_0$ is the radius and $\sigma$ encodes the effective sharpness of the discs' walls. The equation of motion associated with the spin-1 system is obtained from Eq.~\eqref{eqn:ham_s} using  $i\hbar\partial\boldsymbol{\Psi}/\partial t=\delta H/\delta\boldsymbol{\Psi}^{\dagger}$, giving
\begin{equation}\label{eqn:sgpe}
    i\hbar\frac{\partial\boldsymbol{\Psi}}{\partial t}=\bigg(\hat{H}_0+g_nn({\bf r})+g_s\langle\hat{S}({\bf r})\rangle\cdot\hat{S}\bigg)\boldsymbol{\Psi}.
\end{equation}
In the homogeneous limit appropriate to Eq.~\eqref{eqn:sgpe}, the ground state in the broken-axisymmetric phase can be parameterised as $\Psi_{\rm BA}=\sqrt{n_0}e^{i\theta}e^{-i\hat{S}_z\varphi}(\cos\alpha\sin\beta,\cos\beta,\sin\alpha\sin\beta)^T$ \cite{murata_2007} where the angles $\theta$ and $\varphi$ are associated with global phase and spin rotations respectively; while the quantities $\alpha$ and $\beta$ are related to $p$ and $q$ by
\begin{subequations}\label{eqn:ab}
\begin{align}
    \cos2\alpha&=\frac{2qp}{q^2+p^2}, \\
    \cos^{2}\beta&=\frac{(q^2-p^2)(p^2+q^2+qq_0)}{2q^3q_0},
\end{align}
\end{subequations}
here $q_0=2|g_s|n_0$ defines the critical point at the edge of the easy-plane phase (viz. Fig.~\ref{fig:gnd}(b)), with $n_0$ the mean mass density. The spin densities can be obtained from $\Psi_{\rm BA}$ and \eqref{eqn:ab} using the definitions $\langle\hat{S}\rangle=\langle\hat{S}_x\rangle\hat{e}_{x}+\langle\hat{S}_y\rangle\hat{e}_{y}$  and $\langle\hat{S}_{\parallel}\rangle=\langle\hat{S}_{z}\rangle$ as 
\begin{subequations}\label{eqn:sdens}
    \begin{align}\label{eqn:den_s}
        \langle\hat{S}\rangle&=n_0\frac{\sqrt{(q^2{-}p^2)((p^2{+}qq_0)^2{-}q^4)}}{q^2q_0}(\cos\varphi,\sin\varphi),\\
        \langle\hat{S}_{\parallel}\rangle&=n_0\frac{p(p^2-q^2+qq_0)}{q^2q_0}.\label{eqn:den_p}
    \end{align}
\end{subequations}
The axial spin density $\langle\hat{S}_{\parallel}\rangle$ is finite for $p,q>0$ and in the absence of a linear Zeeman field the spin density Eq.~\eqref{eqn:den_s} reduces to
\begin{equation}
    \langle\hat{S}\rangle=n_0\sqrt{1-q^2/q_{0}^{2}}(\cos\varphi,\sin\varphi).
\end{equation}
Magnetic ground-state phase diagrams are depicted in Fig.~\ref{fig:gnd}. Here panel (a) depicts the antiferromagnetic situation corresponding to $g_s>0$ while the ferromagnetic case is shown in (b) corresponding to $\Psi_{\rm BA}$ and Eq.~\eqref{eqn:ab}. In panel (a) the solid and dashed green lines at $p/|g_s|n_0=\pm1$ enclose a region where the $m=0$ state is unoccupied and the $m=\pm1$ states are immiscible. The corresponding ferromagnetic case for $g_s<0$ is presented in (b). Here the three spin states are miscible within the blue shaded region, while outside the dashed lines $|p|=q$ and $|p|=\sqrt{q(q-q_0)}$, so that individual hyperfine states are individually occupied. 
Figure \ref{fig:gnd}(c) and (d) show examples of how the Zeeman field changes with the spin densities $\langle\hat{S}\rangle$ (Eq.~\ref{eqn:den_s}) and $\langle\hat{S}_{\parallel}\rangle$ (Eq.~\ref{eqn:den_p}) respectively. For fixed finite $p$, the corresponding spin $\langle\hat{S}\rangle$ and axial density $\langle\hat{S}_{\parallel}\rangle$ exist on the interval $\big\{q_{\rm min},q_{\rm max}\big\}=\big\{p,|g_s|n_0+\sqrt{g_{s}^{2}n_{0}^{2}+p^2}\big\}$.
\subsection{Polar Core Vortices}
We are interested in polar-core vortices which are supported in the easy-plane phase with ferromagnetic ($g_s<0$) interactions. The phase winding of the two ferromagnetic states $|m_{\rm F}=\pm1\rangle$ are opposite, and for a winding number $\kappa\in\mathds{Z}$ the general state of an individual PCV centred at the origin can be written as 
\begin{equation}\label{eqn:spinv}
    \Psi_{\rm PCV}({\bf r})\approx\sqrt{n_0}
    \left(\begin{array}{c}
        \cos\alpha({\bf r})\sin\beta({\bf r}) e^{-i\kappa\phi({\bf r})} \\
        \cos\beta({\bf r})\\
        \sin\alpha({\bf r})\sin\beta({\bf r}) e^{i\kappa\phi({\bf r})} 
    \end{array}\right)
\end{equation}
here $\phi({\bf r})=\operatorname{arg}[x+iy]$ and the quantities $\alpha({\bf r})$ and $\beta({\bf r})$ are generally spatially varying due to the requirement that the densities present in the $m_{\rm F}=\pm1$ components vanish at the centre of rotation. 

\section{\label{sec:num}Numerical results}
In this section we explore the polar core vortex solutions to the spin-1 Gross-Pitaevskii model, Eq.~\eqref{eqn:sgpe} using a split-operator Crank-Nicolson method (see Appendix \ref{sec:app} for details). We work in a general set of dimensionless units such that the length is scaled in terms of $\xi$, the energy in terms of $\hbar^2/m\xi^2$ so that the density-density and spin interactions $g_{n,s}$ become $g_{n,s}\rightarrow \overline{g}_{n,s}=M\xi^2g_{n,s}n_0/\hbar^2$.

We can make an estimate of the physical size and atom numbers of spinor vortex states on the basis of existing work. The size of a spin excitation is characterized by the spin healing length $\xi_s=\hbar/\sqrt{Mq_0}$, where for typical densities $n_{\rm 3D}\approx 10^{13}~\text{cm}^{-3}$ and scattering lengths appropriate to $^{87}$Rb \cite{vankempen_2002} one has $q_0/h\approx 10$~Hz. This gives $\xi_s\approx3$~$\mu$m while the corresponding dimensionless density $\overline{n}_{3D}=\ell_z\xi_{s}^{2}n_{\rm 3D}$ with $\ell_z=10$~$\mu$m, giving a corresponding atom number $N=\ell_z\pi R_{0}^{2}n_{\rm 3D}$. Then one obtains an approximate atom number on the disc $N\approx20\times 10^{6}$ corresponding to $n_{\rm 2D}\approx 3\times 10^{6}$~\text{cm}$^{-2}$. Then in all simulations we use $\overline{g}_n=10^3$ and $\overline{g}_{s}=-10$ such that $|\overline{g}_s/\overline{g}_n|=0.01$ similar to the value for $^{87}$Rb. For the disc potential Eq.~\eqref{eqn:disc} we take $m\xi^2U_0/\hbar^2=500$, $\sigma/\xi=1/4$ and $R_0/\xi=0.9L_x$. The radius of the disc is taken to be $R_0=50$~$\mu$m~\cite{gauthier_2016}.       
\begin{figure}[t!]
    \centering\hspace{-1.75cm}
    \includegraphics[width=1.0\columnwidth]{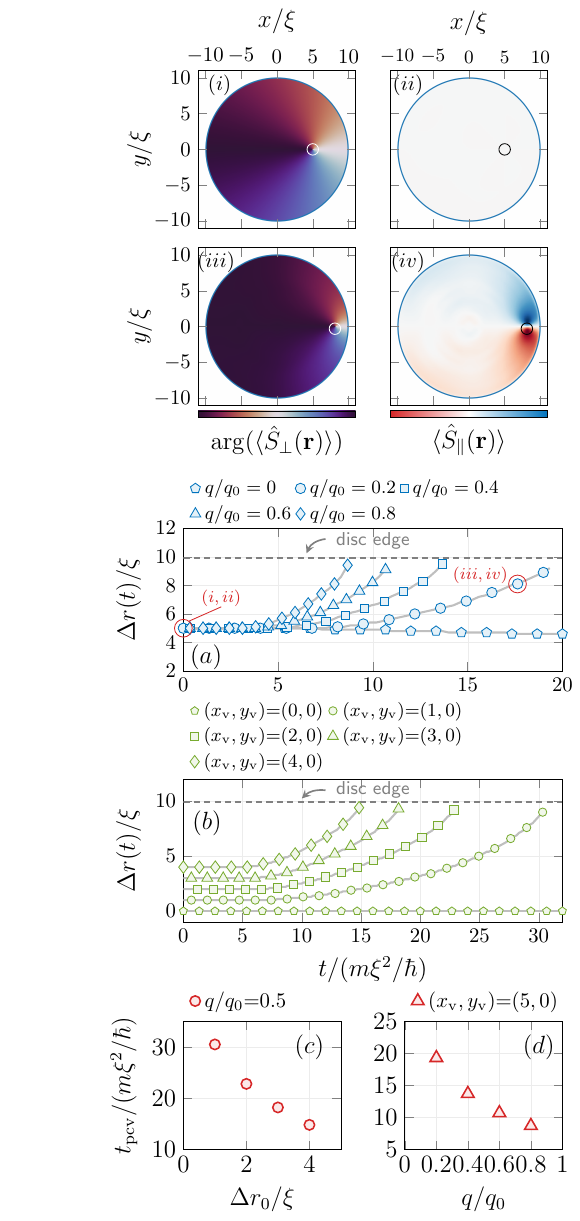}
    \caption{(color online) Single PCV dynamics. Panels (a) and (b) explore the displacement of a PCV as a function of time, while (c) and (d) show the corresponding lifetime $t_{\rm pcv}$ of a vortex. Example initial states corresponding to (a) are presented in (i) and (ii), showing the phase of the transverse  $\langle\hat{S}_{\perp}({\bf r})\rangle$ and axial spin densities $\langle\hat{S}_{\parallel}({\bf r})\rangle$ respectively.}
    \label{fig:sngl}
\end{figure}
\subsection{Individual Polar Core Vortices}
To build intuition on the dynamics of a single polar core vortex confined on a circular disc, we scrutinize the dynamics of a single spin excitation under two sets of initial conditions. The state of the single polar core vortex can be obtained from Eqs.~\eqref{eqn:ab} and Eq.~\eqref{eqn:spinv} for a particular choice of $q$ with $p=0$.  

The dynamics of a single polar core vortex is explored in Fig.~\ref{fig:sngl}. An example initial state for $r_0 = 5\xi$ is plotted in Fig.~\ref{fig:sngl}(i) and (ii) showing the initial phase of the transverse spin $\langle\hat{S}_{\perp}({\bf r})\rangle=\langle\hat{S}_x({\bf r})\rangle+i\langle\hat{S}_y({\bf r})\rangle$ and axial spin density $\langle\hat{S}_{\parallel}({\bf r})\rangle$ respectively. Figure~\ref{fig:sngl}(iii) and (iv) show the state at a later time as indicated by the red labels in (a). The position of the vortex is highlighted by a circle in each case and moves radially to the condensate boundary.
\begin{figure*}[t!]
    \centering
    \includegraphics[width=\linewidth]{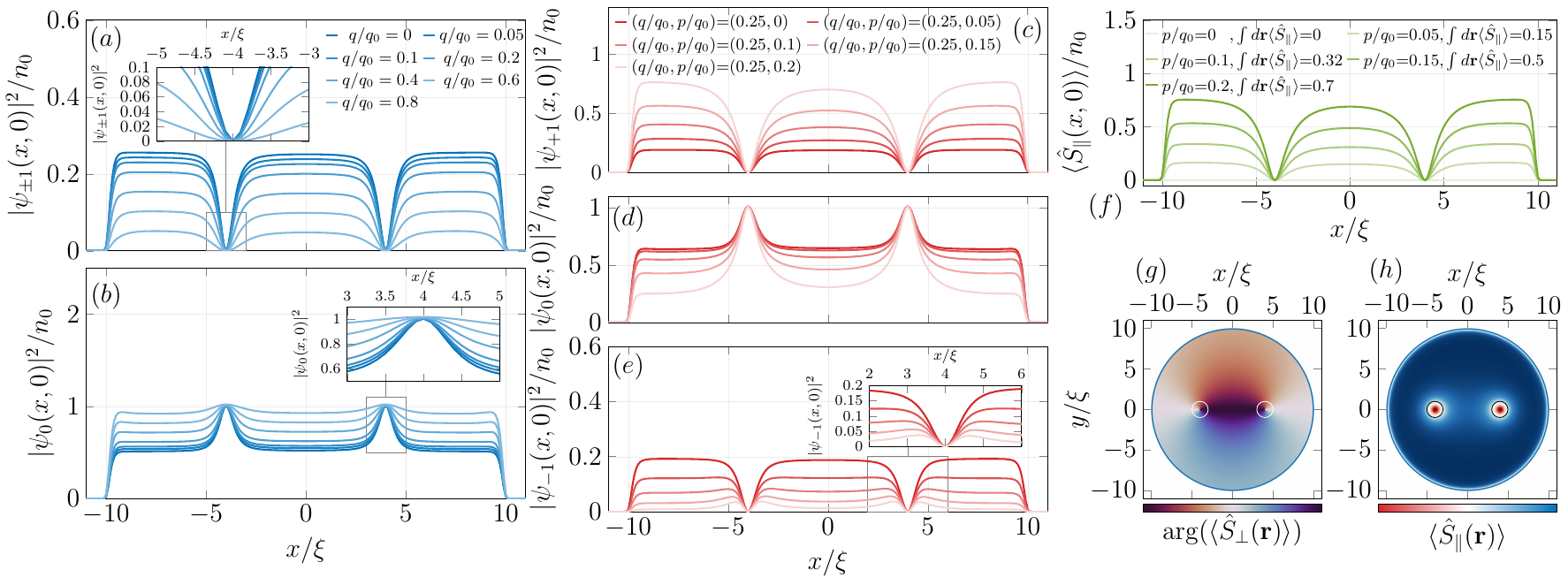}
    \caption{(color online) PCV pair stationary states. Stationary solutions to Eq.~\eqref{eqn:sgpe} with the initial state Eq.~\eqref{eqn:mpcv} are presented. (a) and (b) show cross-sections of the hyperfine densities $|\psi_{m}({\bf r})|^2$ for $p/q_0=0$. Then panels (c)-(e) explore the polar core vortex densities with fixed $q/q_0=0.25$ for $p/q_0>0$. Panel (f) shows the axial spin densities $\langle\hat{S}_{\parallel}({\bf r})\rangle$ corresponding to the data in (c)-(e). The final two panels, (g) and (h) depict the phase of the transverse $\langle\hat{S}_{\perp}({\bf r})\rangle$ and axial $\langle\hat{S}_{\parallel}({\bf r})\rangle$ spin densities for $(q/q_0,p/q_0)=(0.25,0.1)$.}
    \label{fig:pair}
\end{figure*}
This behaviour can be understood by considering the interaction between the vortex and background currents. The particle flow for all three components must point azimuthally at the condensate boundary, as particles do not flow across the boundary \cite{fetter_2009}. For a scalar condensate of radius $R_0$ containing a single vortex at position $\mathbf{r}_0$, this boundary condition is achieved by including an oppositely charged image vortex at position~\cite{groszek_2018,pointin_1976,viecelli_1995}
\begin{equation}\label{eq:image}
    \mathbf{r}_\mathrm{image}=\frac{R_0^2\mathbf{r}_0}{|\mathbf{r}_0|^2}.
\end{equation}
For a spin-1 condensate containing a polar core vortex image charges for both spin components must be included. The polar core vortex dynamics is then described by its attraction to its oppositely charged image vortex~\cite{turner_2009,williamson_2016,williamson_2021}, resulting in outward radial acceleration.

In Fig.~\ref{fig:sngl}(a), the displacement of the PCV from the disc centre is shown $\Delta r(t)=\sqrt{x_{0}(t)^2+y_{0}(t)^2}$ as a function of time for a fixed initial displacement ${\bf r}_0=(5\xi,0)$ for different values of the quadratic Zeeman strength $q/q_0$. As this term is increased, we can see that the dynamics of the excitation change -- for increasing $q/q_0$, the time the vortex takes to reach the edge of the disc and annihilate decreases, consistent with results in a planar system~\cite{williamson_2021}. Physically we can interpret this result in terms of the effect changing $q/q_0$ has on the hyperfine states. For $q\rightarrow 0$, the polar $|m_{\rm F}=0\rangle$ component fills the other hyperfine states such that the effective mass of the filled components is a maximum. Then, as the Zeeman field strength is increased, the $|m_{\rm F}=\pm1\rangle$ states begin to reduce in population. The depth of the cores of the vortices in these components is reduced, and so is the filling from the $|m_{\rm F}=0\rangle$ component, and hence the polar core vortices dynamics follow a shorter timescale, as shown in Fig.~\ref{fig:sngl}(a). The decrease in inertia with increasing $q/q_0$ can also be argued from a microscopic model of the interaction between a polar core vortex and background flow fields~\cite{stevenshough_2024}.

The dynamics of the PCVs can also be interpreted in terms of an analogy with electrodynamics --- here the polar core vortex can be thought of as a charge in an effective electric field \cite{neu_1990}. The dynamics corresponding to $\Delta r(t)/\xi$ are also shown for $q/q_0=0$, pentagons in panel (a). Here the PCVs' dynamics differ to those at finite $q/q_0$, appearing to drift away from the edge of the disc, while over a longer integration time the PCV does not reach the edge of disc, nor approach the center, oscillating instead close to its initial position, attributed to perturbations resulting from density excitations that develop on the surface of the disc at long times. 

Next in Fig.~\ref{fig:sngl}(b) the effect of varying the  initial position of the PCV ${\bf r}_0$ is explored for fixed $q/q_0=1/2$. When the excitation is initially placed at the origin ${\bf r}_0=(0,0)$, the symmetry of the initial condition results in the vortex remaining at its initial position, in contrast to the off-axis example shown in \ref{fig:sngl}(a). Then, as the initial displacement of the excitation is moved away from the origin, one can see the timescale of the dynamics reduces, with the vortex accelerating towards the boundary of the disc as ${\bf r}_0$ increases. The final two panels Fig.~\ref{fig:sngl}(c) and (d) show the lifetime $t_{\rm pcv}$ of the polar core vortex on the disc, extracted from panels (a) and (b). In both cases, increasing either the initial displacement $\Delta r_0$ or quadratic Zeeman strength $q/q_0$ causes $t_{\rm pcv}$ to decrease. 
\begin{figure}[t!]
    \centering
    \includegraphics[width=1\columnwidth]{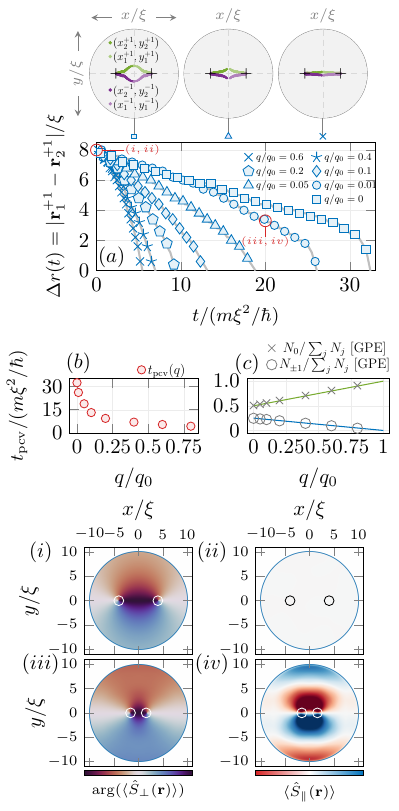}
    \caption{(color online) Polar core vortex dipole dynamics. Panel (a) shows the displacement $\Delta r(t)$ of vortex dipoles for different $q/q_0$, with individual (x(t),y(t)) trajectories shown above for three different values of $q/q_0$. Panel (b) depicts the lifetimes $t_{\rm pcv}$ obtained from (a), while (c) compares the numerical hyperfine populations $N_j$ with their analytical counterparts, $|\Psi_{\rm BA}|^2$ and Eq.~\eqref{eqn:ab}. Examples of the transverse $\langle\hat{S}_{\perp}({\bf r})\rangle$ and axial $\langle\hat{S}_{\parallel}({\bf r})\rangle$ spin dynamics, corresponding to the red markers in (a) are shown in panels (i)-(iv).}
    \label{fig:dipcv}
\end{figure}
\subsection{\label{sec:pcvp}Polar core vortex pairs}
Next we examine the properties of pairs of polar core vortices confined on the circular disc. We consider a general initial condition for the state of the excitation, generalizing Eq.~\eqref{eqn:spinv} such that \cite{williamson_2021}
\begin{equation}\label{eqn:mpcv}
    \Psi_{\rm i}({\bf r}){\approx}\sqrt{n_0}\left(\begin{array}{c}
         \cos\alpha\sin\beta\prod_{j}g_{1}^{j}({\bf r})e^{-i\kappa_j\phi({\bf r})}  \\
         \cos\beta\prod_j g_{0}^{j}({\bf r})  \\
         \sin\alpha\sin\beta\prod_j g_{-1}^{j}({\bf r})e^{i\kappa_j\phi({\bf r})}  
    \end{array}\right)
\end{equation}
and the $g_{j}({\bf r})$ account for the spatial structure of an individual vortex solution, having the properties $g_{\pm 1}({\bf 0})=0$, $g_{0}({\bf 0})\cos\beta\approx 1$ and $g_{j}({\bf r})\approx 1$ when $|{\bf r}|>\xi_s$ \cite{williamson_2016}, while Eq.~\eqref{eqn:mpcv} also allows for a finite initial axial magnetization $\langle\hat{S}_{\parallel}({\bf r})\rangle$.

We first explore the allowed nonlinear stationary states with the ${\bf r}_j$ pinned corresponding to Eq.~\eqref{eqn:mpcv}, depicted in Fig.~\ref{fig:pair}. Here we take $\kappa_1=+1$, $\kappa_2=-1$, ${\bf r}_1=(4\xi,0)$ and ${\bf r}_2=(-4\xi,0)$. Panels \ref{fig:pair}(a) and (b) show cross-sections of the two-dimensional density $|\psi_m(x,0)|^2$ for $p/q_0=0$, with the vortex carrying states $|\psi_{\pm 1}(x,0)|^2$ shown in Fig.~\ref{fig:pair}(a), while the density of the polar state $|\psi_0(x,0)|^2$ is shown in (b). The effect of increasing $q/q_0$ can be clearly seen -- for $q/q_0=0$, the hyperfine populations $\int d{\bf r}\ n_{\pm 1}({\bf r})=\sfrac{1}{4}$ and $\int d{\bf r}\ n_{0}({\bf r})=\sfrac{1}{2}$ results in a \textit{maximum} background density for the $|m_{\rm F}=\pm 1\rangle$ and a \textit{minimum} background density for the $|m_{\rm F}=0\rangle$. Then, as $q/q_0$ increases (shaded blue data) the two vortex carrying components' background densities gradually decrease, while the filled polar component $|m_{\rm F}=0\rangle$ instead begins to increase. The two insets show an enlarged region of the spatial structure of the vortex core / filled component.

Then we explore the effect of having a finite linear Zeeman field $p/q_0\neq 0$, shown in Fig.~\ref{fig:pair}.(c)-(e). Cross sections of the individual hyperfine densities $|\psi_m(x,0)|^2$ are shown for a fixed value of $q/q_0=\sfrac{1}{4}$ and $p/q_0=\{0,0.05,0.1,0.15,0.2\}$. Increasing $p/q_0$ has the effect of gradually increasing the population of the $|m_{\rm F}=+1\rangle$ hyperfine state, while reducing those of the $|m_{\rm F}=0,-1\rangle$ states. The result of having a finite linear Zeeman term breaks the symmetry of the $|m_{\rm F}=\pm 1\rangle$ states shown in Figs.~\ref{fig:pair}(a) for $p/q_0=0$. The inset in panel (e) shows how the shape of the vortex cores change as $p/q_0$ is increased. Panel (f) shows cross-sections of the axial spin density $\langle\hat{S}_{\parallel}({\bf r})\rangle$ using the data shown in (c)-(e), and numerical values of the magnetization $\int d{\bf r}\ \langle\hat{S}_{\parallel}({\bf r})\rangle$ are also given. Example transverse and axial spin densities corresponding to $(q/q_,p/q_0)=(0.25,0.1)$ are depicted in Figs.~\eqref{fig:pair}(g) and (h) respectively. Here the presence of the polar core vortex can be clearly seen in both of the spin densities.

\subsection{Polar core vortex dipole dynamics}
Here we explore the dynamics of the polar core vortices discussed in Sec.~\ref{sec:pcvp}. Since each of the $|m_{\rm F}=\pm1\rangle$ hyperfine states contains one $\kappa_1=+1$ and one $\kappa_2=-1$ vortex charge, this situation is analogous to a dipole configuration which has been extensively studied in single-component condensate systems \cite{neely_2010,aioi_2011}. In general the interaction with both the second vortex and the image charges play a role in the dynamics, however for two vortices sufficiently far from the condensate boundary the former effect dominates. This case is explored in Fig.~\ref{fig:dipcv} for two oppositely charged polar-core vortices confined on the circular disc potential Eq.~\eqref{eqn:disc}.

Figure~\ref{fig:dipcv}(a) shows the displacement of the vortices for varying $q/q_0$. Here the effect of changing the strength of the quadratic Zeeman field is  explored and the displacement of the vortices
\begin{equation}\label{eqn:rt}
    \Delta r^{+1}(t){=}\sqrt{\big(x_{1}^{+1}(t){-}x_{2}^{+1}(t)\big)^2{+}\big(y_{1}^{+1}(t){-}y_{2}^{+1}(t)\big)^2}
\end{equation}
is calculated for varying $q/q_0$. Increasing the effective Zeeman field strength $q/q_0$ has the effect of decreasing (accelerating) the timescale of the vortices dynamics which move towards each other before annihilating. Individual trajectories $\big(x_{k}^{\pm1}(t),y_{k}^{\pm1}(t)\big)$ are presented above panel (a) corresponding to the data for $q/q_0=0$ (square), $q/q_0=0.05$ (triangle) and $q/q_0=0.6$ (cross). The effect of the Zeeman term on the dynamics can be clearly seen -- when $q/q_0=0$ the intercomponent vortices have a pronounced transverse component to their trajectory on the disc (termed ``stretching'' in~\cite{turner_2009,williamson_2016,williamson_2021}). Then as $q/q_0$ is increased this effect is diminished such that for large $q/q_0$ the intercomponent vortices almost directly attract each other travelling in a straight line. The lifetime of the vortices is quite sensitive to $q/q_0$, falling off rapidly. Then, panel (b) presents the lifetime $t_{\rm pcv}$ computed from the data in (a), showing a rapid decay as $q/q_0$ increases. Then panel (c) presents a comparison of the individual hyperfine populations $N_m=\int d{\bf r}|\psi_m({\bf r})|^2$ computed numerically (cross and circle markers) with the analytical prediction obtained from $|\Psi_{\rm BA}|^2$ and Eq.~\eqref{eqn:ab} with $\alpha=\pi/4$ showing close agreement. Finally examples of the vortices dynamics from panel (a) are presented in (i)-(iv). Here the phase of the transverse $\langle\hat{S}_{\perp}({\bf r})\rangle$ and axial spin densities $\langle\hat{S}_{\parallel}({\bf r})\rangle$ are shown for the initial stationary state at $t=0$ for $q/q_0=0.01$ in (i) and (ii), and at $\hbar t/m\xi^2\approx 20$ for panels (iii) and (iv). The circles indicate the location of the individual polar core vortices in the density.

\begin{figure}[b!]
    \centering
    \includegraphics[width=1\columnwidth]{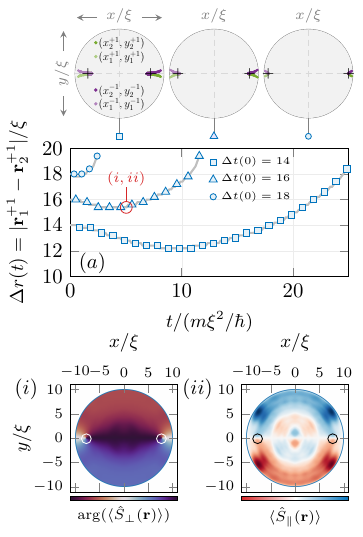}
    \caption{\label{fig:edge}(color online) Vortex dipole disc edge dynamics. (a) shows the displacement of PCV dipoles at large initial separations, corresponding trajectories shown per the grey discs. (i) and (ii) present examples of the spin densities ${\rm arg}(\langle\hat{S}_{\perp}({\bf r})\rangle)$ and $\langle\hat{S}_{\parallel}({\bf r})\rangle$ per the label in (a)}
\end{figure}

The finite size of the disc potential, Eq.~\eqref{eqn:disc} facilitates an opportunity to explore the dynamics of the excitations in proximity to the boundary. Figure \ref{fig:edge} presents simulations exploring the dynamics where PCV dipoles are placed initially at $(x,y)=(\pm n\xi,0)$ where $n=7,8,9$ and $q/q_0=0.1$. Polar core vortex pairs are found to begin to move together (see local minima in panel (a)) before moving towards the disc edge at longer times and annihilating. Examples of the excitations' dynamics are given in panels (i) and (ii) showing the transverse ${\rm arg}(\langle\hat{S}_{\perp}({\bf r})\rangle)$ and parallel $\langle\hat{S}_{\parallel}({\bf r})\rangle$ spin densities respectively for the data point indicated in (a). As the initial position of the vortices are moved closer to the boundary the dynamics qualitatively changes, as the interaction with the image charges dominates the dynamics. In this regime the pair of oppositely charged vortices move radially outward, see Fig.~\ref{fig:edge}. We can estimate the cross-over in behaviour by equating the distance separating the dipole to the distance between the real and image vortices, $2|\mathbf{r}_0|=|\mathbf{r}_\mathrm{image}-\mathbf{r}_0|$ [see Eq.~\eqref{eq:image}], giving $|\mathbf{r}_0|=R_0/\sqrt{3}\approx 6\xi$. This is close to the cross-over point we identify numerically, with the small difference likely arising from the softness of the trap boundary.
\begin{figure}[t!]
    \centering
    \includegraphics[width=0.9\columnwidth]{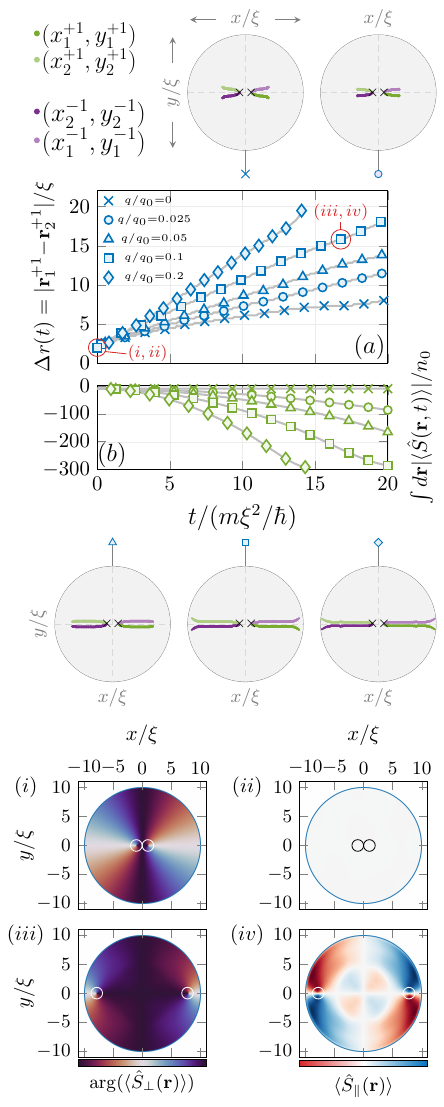}
    \caption{(color online) Same-sign polar core vortex dynamics. Panel (a) shows the displacement as a function of time $\Delta r(t)$ of vortex pairs for different values of $q/q_0$, while (b) computes the corresponding spin density $\int d{\bf r}\langle\hat{S}({\bf r},t)\rangle/n_0$. Trajectory data in (a) are depicted in individual discs for $|m_{\rm F}=\pm1\rangle$ hyperfine states. (i)-(iv) show example spin dynamics of $\langle\hat{S}_{\perp}({\bf r})\rangle$ and $\langle\hat{S}_{\parallel}({\bf r})\rangle$ taken from (a).}
    \label{fig:sspcv}
\end{figure}
\subsection{Same-sign PCV dynamics}
Next we consider the dynamics of same-sign polar core vortices, which has not been explored even in planar systems. The corresponding initial state for the system is obtained from Eq.~\eqref{eqn:mpcv} with $\kappa_1=1$, $\kappa_2=1$, ${\bf r}_1=(\xi,0)$, ${\bf r}_2=(-\xi,0)$ and a total integration time $t=20m\xi^2/\hbar$. Figure \ref{fig:sspcv}(a) shows the displacement $|{\bf r}_{1}^{+1}-{\bf r}_{2}^{+1}|/\xi$ of the same-sign vortex pairs. The dynamics of the polar core vortices are observed to be repulsive, with the time scale of their dynamics reducing as the Zeeman strength $q/q_0$ increases. Individual trajectory plots are presented on the accompanying grey discs corresponding to the data in (a). In analogy with the dynamics observed in Fig.~\ref{fig:dipcv}, for low values of $q/q_0$ the vortices trajectories display a transverse component along $y$ in the $(x,y)$ plane which is reduced as $q/q_0$ increases, eventually leading to quasi-linear dynamics, except close to the disc edge at later times where the vortex trajectory becomes transverse to the disc boundary (square and diamond data). Panel (b) shows the averaged spin density for the data presented in (a) as a function of time, here in general $\int d{\bf r}|\langle\hat{S}({\bf r},t)\rangle|/n_0$ decreases for finite $q/q_0$ as a function of time, while the spin density is independent of time for $q/q_0=0$. Examples of the polar core vortices dynamics, for $q/q_0=0.1$ are shown in panels (i)-(iv). The position of the vortices' cores are highlighted (circles) in all cases. The axial spin density $\langle\hat{S}_{\parallel}({\bf r})\rangle$ displays the formation of Chladni-like patterns \cite{mateo_2014} in (iv) attributed to the underlying circular geometry of the disc potential.  
\begin{figure*}[t!]
    \centering
    \includegraphics[width=0.9\linewidth]{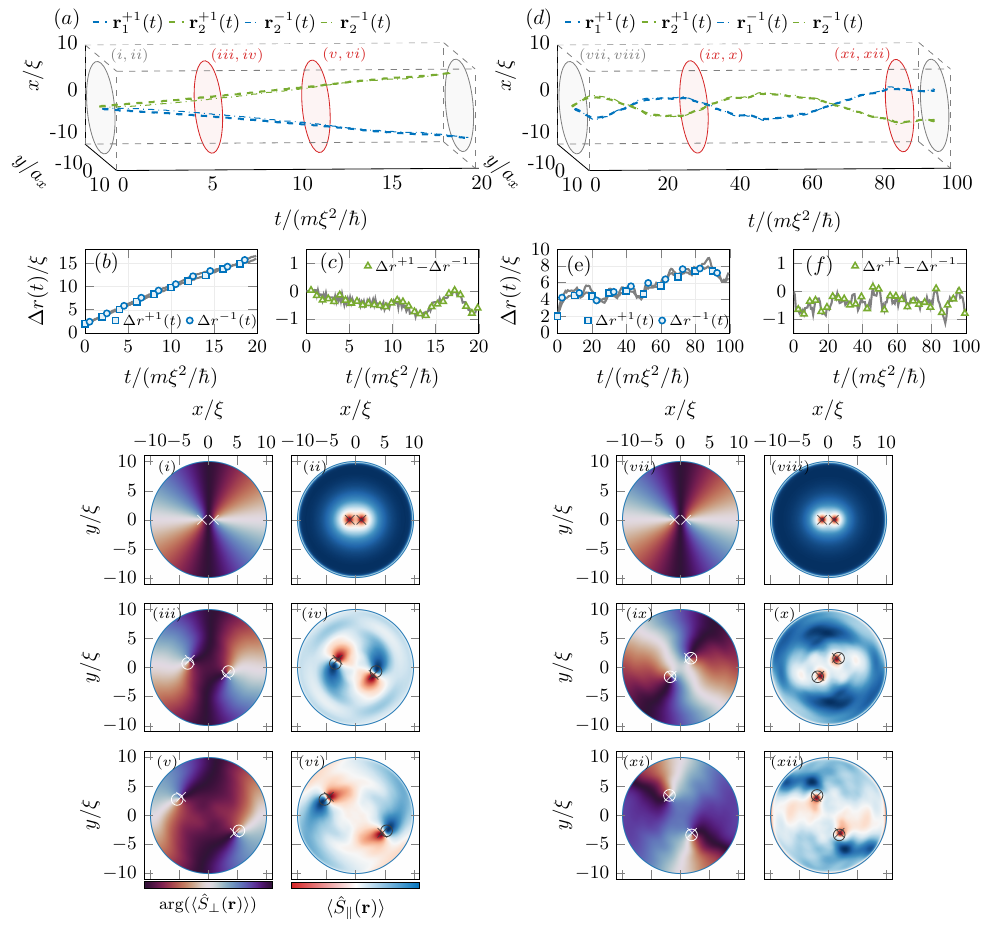}
    \caption{(color online) Finite axial magnetization dynamics. Individual vortex trajectories ${\bf r}_{k}^{\pm1}(t)/\xi$ are shown in (a) and (d), while displacement data $\Delta r^{\pm 1}(t)$ are depicted in (b)-(c) and (d)-(f). Example vortex spin densities $\text{arg}(\langle\hat{S}_{\perp}({\bf r})\rangle)/n_0$ and $\langle\hat{S}_{\parallel}({\bf r})\rangle/n_0$ are shown for the times $\hbar t/m\xi^2\approx0,6,12$ (i)-(vi) and $\hbar t/m\xi^2\approx0,33,90$ (vii)-(xii).}
    \label{fig:spiral}
\end{figure*}
\subsection{PCV dynamics with finite axial magnetization}
So far we have restricted our analysis to zero net axial magnetization in which case the pairs of polar core vortices in Fig.~\ref{fig:sspcv} move radially outward. In contrast, same-sign vortices in a scalar condensate with the same initial setup would circulate the trap centre~\cite{navarro_2013,murray_2016,groszek_2018}. By changing the net axial magnetization, pairs of polar core vortices can be tuned between radial motion (no net axial magnetization) and azimuthal motion (fully axially magnetized). 

To further quantify the polar core vortex phenomenology explored in Fig.~\ref{fig:spiral}, we examine the dynamics of pairs of same-sign polar core vortices with the initial positions ${\bf r}_1=(+\xi,0)$ ($\kappa_1=+1$) and ${\bf r}_2=(-\xi,0)$ ($\kappa_2=+1$) but with a finite initial axial magnetization $\langle\hat{S}_{\parallel}({\bf r})\rangle$ such that $(q/q_0,p/q_0)=(0.1,0.01)$ corresponding to panels (a)-(c) and (i)-(vi) while $(q/q_0,p/q_0)=(0.2,0.1)$ corresponds instead to panels (d)-(f) and (vii)-(xii). The magnetizations in each case are $\int d{\bf r}\langle\hat{S}_{\parallel}({\bf r})\rangle/n_0\simeq0.1,0.4$ respectively. The trajectories ${\bf r}_{k}^{\pm 1}(t)$ of the vortices in the $|m_{\rm F}=\pm 1\rangle$ hyperfine components are shown in (a) corresponding to $(q/q_0,p/q_0)=(0.1,0.01)$. The repulsive dynamics as explored previously in Fig.~\ref{fig:sspcv} for $p/q_0=0$ are evident, however due to the finite linear Zeeman term there is a `twist' beginning to occur. The relative displacements of the vortices 
\begin{equation}\label{eqn:rt2}
    \Delta r^{\pm 1}(t){=}\sqrt{\big(x_{1}^{\pm 1}(t){-}x_{2}^{\pm 1}(t)\big)^2{+}\big(y_{1}^{\pm 1}(t){-}y_{2}^{\pm 1}(t)\big)^2}
\end{equation}
where ${\bf r}_{k}^{\pm 1}=(x_{k}^{\pm 1},y_{k}^{\pm 1})$ is depicted in (b), showing the linear growth of the polar core vortices displacements. Panel (c) shows the difference of the hyperfine displacements $\Delta r^{+1}(t)-\Delta r^{-1}(t)$ (green data). The centers of rotation of the vortices are no longer the same due to the finite Zeeman term, leading to a slight difference in the vortices displacements in the different hyperfine states. Then, examples of the dynamics corresponding to panels (a)-(c) are shown in (i)-(vi). Each of the two columns shows the phase of the transverse spin density $\text{arg}(\langle\hat{S}_{\perp}({\bf r})\rangle)/n_0$ (left) and axial spin density $\langle\hat{S}_{\parallel}({\bf r})\rangle/n_0$ (right). The centers of rotation for individual vortices have been tracked and are shown as cross and circle markers for $\hbar t/m\xi^2\approx6$ (iii,iv) and $\hbar t/m\xi^2\approx 12$ (v,vi). Both time-dependent spin densities exhibit the twisting effect of the vortices dynamics. The example dynamics presented in (i)-(vi) are highlighted in (a) at the appropriate times as annotated coloured discs. The trajectories ${\bf r}_{k}(t)$ of the polar core vortices for $(q/q_0,p/q_0)=(0.2,0.1)$ are presented in (d). The increased strength of both contributions to the Zeeman energy result in spiral dynamics where the same-sign pair simultaneously repel and co-rotate. Again the corresponding relative displacements $\Delta r^{\pm 1} (t)/\xi$ are shown in (e) for both hyperfine states (blue data) while the difference $\Delta r^{+1} (t)-\Delta r^{-1} (t)$ (green data) is shown in panel (f). The dynamics of the spin densities displayed in (vii)-(xii) show the interaction of the phase defects, here the combination of the repulsion and revolution of the vortices centers of rotation is observed for $\hbar t/m\xi^2\approx33,90$ (see corresponding highlighted discs in (d)). We note that related work also studied topological vortex dynamics in spin systems exhibiting attractive spiral dynamics \cite{chojnacki_2024}. 

In order to further understand and quantify the spiral dynamics of the polar core vortices presented in Fig.~\ref{fig:spiral}(d)-(f), we make a heuristic comparison between the numerical vortex positions ${\bf r}_{k}^{\pm 1}(t)$ and an analytical approximation given by an Archimedes-like spiral 
\begin{equation}\label{eqn:as}
    {\bf r}_{k}^{\rm A}(t)=\big(ut+x_{k}\big)\left(\begin{array}{c}
         \cos(\omega_k(t) t)  \\
         -\sin(\omega_k(t) t)
    \end{array}\right).
\end{equation}
Here the effective radial velocity is $u$ while $x_{k}$ is the initial position of vortex $k$, accounting for the repulsive component of the spinor vortex dynamics, while $\omega_k(t)=\mathcal{C}/(ut+x_{k})$ defines the time-dependent frequency of vortex $k$. As the vortices separate, the flow field experienced by each vortex due to the other vortex decreases \cite{ueda_book}, which gives rise to a decrease in $\omega_k$. We note, however, that our heuristic form $\omega_k(t)\sim1/|r|$ differs from that of vortices in scalar condensates (which follows $\omega\sim1/r^2$, neglecting effects of the trapping potential) \cite{navarro_2013}. We attribute this to the spinor nature of the superfluid and effects of the trap.

\begin{figure}[t!]
    \centering\hspace{-0.5cm}
    \includegraphics[width=1.0\columnwidth]{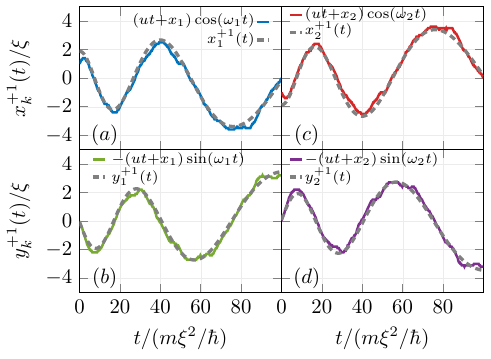}
    \caption{(color online) Vortex spiral comparison. Coloured data showing ${\bf r}_{k}(t)$ taken from Fig.~\ref{fig:spiral}(d), with the fit (dashed grey lines) computed from Eq.~\eqref{eqn:as}.}
    \label{fig:sfit}
\end{figure}

Comparisons between Eq.~\eqref{eqn:as} and the corresponding numerical data taken from Fig.~\ref{fig:spiral}(d) are presented in Fig.~\ref{fig:sfit}. Panels (a) and (c) here show the $x_{k}^{+1}(t)/\xi$ data while (b) and (d) show the $y_{k}^{+1}(t)/\xi$ comparison. Coloured data corresponds to the numerical vortex trajectories while the dashed grey curves show the fits obtained from Eq.~\ref{eqn:as}. Here we find $um\xi/\hbar\approx0.018$, $x_{1,2}/\xi\approx\pm1.83$ and $m\mathcal{C}\xi/\hbar\approx0.4$.  

The heuristic model Eq.~\eqref{eqn:as} is well supported by the numerical data since the time-dependent frequency $\omega_k(t)$ enables us to capture the dynamics of the vortices as they spiral outwards. Slight differences with the numerical trajectories computed from the spinor Gross-Pitaevskii model Eq.~\ref{eqn:sgpe} are attributed here to the full nonlinear dynamics that the modest heuristic model Eq.~\eqref{eqn:as} does not capture.   
\section{Summary}
In this work the nonlinear dynamics of single and pairs of polar core vortices confined on a circular disc in an atomic spin-1 condensate were explored. Numerical solutions to the spin-1 Gross-Pitaevskii model revealed the unusual interplay of the confining geometry and the excitations. In contrast to scalar condensates, a single PCV confined on the disc and initially displaced from the centre will in general accelerate towards the edge of the confining geometry before annihilating at the potential edge. This behaviour  was interpreted by considering an oppositely charged image PCV that attracts the excitation towards the disc's edge.

The dynamical behaviour of oppositely charged (dipole) PCVs were studied, showing two distinct behaviours. For vortex pairs placed initially close together, the excitations dynamics were found to be attractive, with the PCVs moving towards each other before annihilating. On the other hand when the excitations were in closer proximity to the edge of the disc, the PCVs begun to move together at short times, before changing direction and annihilating with the edge of the confining potential.

For the case of same-sign PCVs, pair dynamics were found to be repulsive generally, with the strength of the repulsion increasing with the Zeeman field strength, again the PCV pairs eventually annihilated with the potential boundary at long times. The effect of an initial axial magnetization was also examined in this context, revealing spiral dynamics where the PCVs simultaneously repel and rotate. The dynamics of the vortex cores were compared to an Archimedes' spiral, showing good agreement. 

It would be interesting to understand this system further, particularly the dynamics of PCV dipoles placed close to the edge of the disc and the origin of the minima observed in Fig.~\ref{fig:edge}(a), as well as probing the robustness of the topological states via interaction quenches. This work opens the door for exploring spinor vortex dynamics in novel homogeneous potentials such as square, triangular or more elaborate geometrical confinements such as annuli \cite{eckel_2014} and lemniscates \cite{bland_2022} which connect to emerging applications in atomtronics.

\section{Acknowledgements}
We thank Guillaume Gauthier and Andrew Groszek for discussions. M.E was supported by the Australian Research Council Centre of Excellence in Future Low-Energy Electronics Technologies (Project No.~CE170100039), and by the Japan Society of Promotion of Science Grant-in-Aid for Scientific Research (KAKENHI Grant No.~JP20K14376) and by an InternKNOTship at the WPI program “Sustainability with Knotted Chiral Meta Matter (SKCM2)” at Hiroshima University. L.A.W. was supported by the Australian government Department of Industry, Science, and Resources via the Australia-India Strategic Research Fund (AIRXIV000025).

\appendix
\section{\label{sec:app}Spinor split-operator method}
The system of equations defined by Eqs.~\eqref{eqn:sgpe} can be decomposed into matrix form as
\begin{equation}\label{eqn:sosplit}
    i\hbar\frac{\partial}{\partial t}\left[\begin{array}{c} \psi_{1} \\ \psi_{0} \\ \psi_{-1} \end{array}\right]=\bigg\{
    \hat{H}_{0}+\hat{H}_{1}+\hat{H}_{2}
    \bigg\}\left[\begin{array}{c} \psi_{1} \\ \psi_{0} \\ \psi_{-1} \end{array}\right],
\end{equation}
where
\begin{subequations}
    \begin{align}&\hat{H}_0=\text{diag}\bigg[\hat{K}-p+q,\hat{K},\hat{K}+p+q\bigg]\label{eqn:h0}
    ,\\ \nonumber
    &\hat{H}_{1}=\text{diag}\bigg[g_nn+g_s\big(n_1+n_0-n_{-1}\big),g_nn+\\&g_s\big(n_{1}+n_{-1}\big),g_nn+g_s\big(-n_{1}+n_{0}+n_{-1}\big)\bigg] \label{eqn:h1}
    ,\\ 
    &\hat{H}_{2}=g_s\left(\begin{array}{ccc}
         0 & \psi_{-1}^{*}\psi_0 & 0 \\
        \psi_{0}^{*}\psi_{-1} & 0 & \psi_{0}^{*}\psi_{1} \\
         0 & \psi_{1}^{*}\psi_{0} & 0 \\
    \end{array}\right).\label{eqn:h2}
    \end{align}
\end{subequations}
Here the operator $\hat{K}=-\frac{\hbar^2}{2M}\nabla^2+U({\bf r})$ is written for brevity. The spinor system is broken down into three constituents -- a single particle term $\hat{H}_0$ (Eq.~\ref{eqn:h0}) which is evolved using Crank-Nicolson, and two mean-field interaction parts (density-density and spin-exchange terms) given by Eqs.~\eqref{eqn:h1} and \eqref{eqn:h2} respectively. 

The contribution to a single time-step from the single-particle term $\hat{H}_0$ on the right-hand side of Eq.~\ref{eqn:sosplit} is obtained from the solution of 
\begin{equation}
    i\hbar\frac{\partial\psi_m}{\partial t}=\hat{H}_0\psi_m,
\end{equation}
for each of the three components $\psi_{+1,0,-1}$, followed by the two nonlinear parts of the time-evolution step. The density-density (spin-conserving) interaction is given by $\Psi^{\rm new}=\mathcal{M}_1\Psi^{\rm old}$ where
\begin{equation}
    \mathcal{M}_1=\text{diag}\big(e^{{-}\frac{idt}{\hbar}\hat{H}_{1,+1}},e^{{-}\frac{idt}{\hbar}\hat{H}_{1,0}},e^{{-}\frac{idt}{\hbar}\hat{H}_{1,-1}}\big),
\end{equation}
here the nonlinear Hamiltonians $\hat{H}_{1,j}$ appearing in $\mathcal{M}_1$ are defined as
\begin{subequations}
\begin{align}
\hat{H}_{1,+1}&=g_nn+g_s(n_1+n_0-n_{-1}),\\
\hat{H}_{1,0}&=g_nn+g_s(n_1+n_{-1}),\\
\hat{H}_{1,-1}&=g_nn+g_s(-n_1+n_{0}+n_{-1}).
\end{align}
\end{subequations}
The second nonlinear contribution to the time-step consists of the exponentiation of $\hat{H}_2$. This step is pre-allocated, and is calculated as $\Psi^{\rm new}=\mathcal{M}_2\Psi^{\rm old}$ where
\begin{equation}\label{eqn:m3}
    \mathcal{M}_2=\mathds{1}_{3\times 3}+\frac{\cos\Omega-1}{\Omega^2}\frac{\Delta t^2}{\hbar^2}(\hat{H}_{2})^{2}-i\frac{\sin\Omega}{\Omega}\frac{\Delta t}{\Omega}\hat{H}_2,
\end{equation}
and $\Omega=g_s\Delta t|\psi_0|\sqrt{|\psi_{-1}|^2+|\psi_{+1}|^2}/\hbar$ \cite{kaur_2021}. The equivalent expression for imaginary time propagation can be obtained from Eq.~\ref{eqn:m3} using the replacement $\Delta t\rightarrow-i\Delta t$.

A single time step of length $\Delta t$ comprises evaluating the single-particle terms for each spin component, followed by applying the two matrix exponential steps, as well as a rescaling step to preserve the total atom number $N$ when working in imaginary time such that
\begin{equation}
    \psi_m({\bf r})\rightarrow\psi_m({\bf r})\bigg\{\sum_m\int d{\bf r}|\psi_{m}({\bf r})|^2\bigg\}^{-1/2}.
\end{equation}
For the pinned stationary-state calculations we took $\hbar\Delta t/(m\xi^2)=10^{-5}$, $\{\Delta x/\xi,\Delta y/\xi\}=0.05$ with a grid size of $(N_x,N_y)=(221,221)$ for individual spin components $\psi_m(x,y)$. Example spin-1 Gross-Pitaevskii Python scripts used to obtain the data is this work are available \cite{github}.


\begin{thebibliography}{99}
\bibitem{pm_def}
The spin-1 matrices are defined as $\hat{S}_x=\frac{\hbar}{\sqrt{2}}\left(\begin{array}{ccc}0 & 1 & 0 \\1 & 0 & 1 \\0 & 1 & 0\end{array}\right)$, $\hat{S}_y=\frac{\hbar}{\sqrt{2}}\left(\begin{array}{ccc}0 & -i & 0 \\i & 0 & -i \\0 & i & 0\end{array}\right)$ and $\hat{S}_z=\hbar\left(\begin{array}{ccc}1 & 0 & 0 \\0 & 0 & 0 \\0 & 0 & -1\end{array}\right)$.
\end{thebibliography}
\end{document}